\documentclass[iop]{emulateapj}

\usepackage{float}
\usepackage{times}
\usepackage{graphicx}
\usepackage{amssymb}
\usepackage{color}

\newcommand{\mbh}{M_\bullet}
\newcommand{\mdef}{M_\mathrm{def}}
\newcommand{\rsoi}{r_\mathrm{soi}}
\newcommand{\rb}{r_b}

\newcommand{\betarot}{\beta_\mathrm{rot}}
\newcommand{\hst}{{\it HST}}

\shorttitle{Core scouring in massive ellipticals}
\shortauthors{Thomas et al.}

\begin{document}

\title{The dynamical fingerprint of core scouring in massive elliptical galaxies}

\author{J. Thomas\altaffilmark{1}, R. P. Saglia,
  R. Bender, P. Erwin, and M. Fabricius} \affil{Max Planck-Institute for
  extraterrestrial Physics, PO Box 1312, Giessenbachstr. 1, 85741
  Garching, Germany\\Universit\"atssternwarte M\"unchen, Scheinerstra\ss e
  1, D-81679 M\"unchen, Germany} \altaffiltext{1}{email: jthomas@mpe.mpg.de}

\begin{abstract}
  The most massive elliptical galaxies have low-density centers or
  cores that differ dramatically from the high-density centers of less
  massive ellipticals and bulges of disk galaxies. These cores have
  been interpreted as the result of mergers of supermassive black hole
  binaries, which depopulate galaxy centers by gravitationally
  slingshotting central stars toward large radii. Such binaries
  naturally form in mergers of luminous galaxies.  Here, we analyze
  the population of central stellar orbits in 11 massive elliptical
  galaxies that we observed with the integral field spectrograph
  SINFONI at the European Southern Observatory Very Large
  Telescope. Our dynamical analysis is orbit-based and includes the
  effects of a central black hole, the mass distribution of the stars
  and a dark matter halo.  We show that the use of integral field
  kinematics and the inclusion of dark matter is important to conclude
  upon the distribution of stellar orbits in galaxy centers.  Six of
  our galaxies are core galaxies. In these six galaxies, but not in
  the galaxies without cores, we detect a coherent lack of stars on
  radial orbits in the core region and a uniform excess of radial
  orbits outside of it: when scaled by the core radius $r_b$, the
  radial profiles of the classical anisotropy parameter $\beta(r)$ are
  nearly identical in core galaxies. Moreover, they match
  quantitatively the predictions of black hole binary simulations,
  providing the first convincing dynamical evidence for core scouring
  in the most massive elliptical galaxies.
\end{abstract}


\section{Introduction}

Elliptical galaxies have long been believed to be the end-product of
collisions, or mergers, between galaxies. Tracing back in time the
evolutionary history of individual ellipticals is difficult, however,
since galaxy collisions involve strong distortions in the
gravitational potential that randomize the distribution of stellar
orbits. Most of the information about the structure of possible
progenitor galaxies and even the number of mergers is therefore
erased.  For example, over several orders of magnitude in radius, the
smooth light profiles of elliptical galaxies can be well described by
the single three-parameter S{\'e}rsic function
\citep{Sersic63,Caon93}.  However, the most massive galaxies often
exhibit surface brightness profiles that are, interior to a ``break''
or ``core'' radius $\rb$, significantly flatter than the inward
extrapolation of the outer, S{\'e}rsic-like profile.
The resulting cores have typical sizes of $\rb \sim 50-500 \,
\mathrm{pc}$. Cores are not the only characteristic by which the most
massive and brightest ellipticals are distinct from fainter elliptical
galaxies: massive ellipticals are less flattened, with boxy instead of
disky isophotes, and are dominated by unordered, rather than
rotational, stellar motions \citep[e.g.][]{Kormendy96,Faber97}. Based on
statistical comparisons between observed galaxies and numerical
$N$-body simulations these morphological differences provide
circumstantial evidence for two distinct formation paths. Fainter
ellipticals likely form from mergers of disk-dominated progenitor
galaxies that are rich in gas; the gas can then settle into a
rotational, disky configuration before forming new stars. The most
massive ellipticals, however, are thought to form through gas-poor
(i.e., elliptical-elliptical) mergers.

The existence of cores represents a challenge to our understanding of
the merging process. The central structure of a merger without gas is
dominated by the more concentrated of the two progenitors: the steeper
central density cusp survives \citep{Fulton01,Boylan-Kolchin04}.
Since high-mass ellipticals are thought to be built from mergers of
lower-mass ellipticals, which have steep central-density cusps, the
shallower cores in more massive ellipticals have to be the result of
another physical mechanism. Supermassive black holes (SMBHs) that
reside in the centers of essentially all elliptical galaxies provide
such a core formation mechanism: during a merger, the two central
SMBHs of the progenitors sink to the center of the remnant by
dynamical friction and form a binary. The binary then loses angular
momentum to stars in the center via three-body interactions, scouring
out the core while becoming progressively more tightly bound until it
eventually merges \citep{Begelman80}.

Although this model currently represents the only convincing scenario
for core formation, only indirect evidence for this process had been
presented previously: the size of the core and the amount of
starlight that is “missing” in the centers of core galaxies have been
found to scale approximately with the mass of the central black hole,
as predicted by theory
\citep{Graham04,KormendyBender09,KormendyHo13}. By their very nature,
however, missing-light measurements depend strongly on model-dependent
assumptions about the original central light profiles
\citep{Hopkins10}.

Core scouring by a black hole binary proceeds after the phase of
violent randomization of stellar orbits by a merger. This is crucial
because stellar orbits then become an important ``archeological''
diagnostic for core formation. The two-body relaxation time for stars
in galaxies exceeds the age of the universe. Galaxies are therefore
collisionless systems and once the stars have settled on their orbits,
the phase-space distribution remains conserved for very long
times. Core scouring changes the orbit distribution: only radial
orbits allow for close passage past the galaxy center and thus only
stars on radial orbits can reach the vicinity of the central binary
black hole and get ejected. Consequently, the orbital structure left
behind in the core after core scouring is predicted to be strongly
biased in favor of tangential orbits, while the ejected stars contribute to enhanced
radial motions outside the core \citep{Quinlan97,Milosavljevic01}.

That stellar orbits around black holes are predominantly tangential is
predicted not only by SMBH-binary models. The traditional picture for
the formation of non-core galaxies through gas-rich mergers leads to
an enhanced population of tangential orbits as well, because the new
stars that form from gas that falls in during/after the merger are
likely to end up in a rotating structure that is supported by
near-circular orbits.  Likewise, the changes in the gravitational
potential induced by an (adiabatically) growing black hole that is fed
by gas reshape stellar orbits toward becoming more tangential as well
\citep{Young80,Goodman84,Quinlan95}.

Previous dynamical studies addressing the distribution of stellar
orbits in the vicinity of SMBHs focused mainly on
the mere detection of tangential orbits, but did not try a direct
comparison with predictions from theoretical models. This is
necessary, however, to distinguish between dissipational and
dissipationless core formation, since both predict stars to
predominantly move along tangential orbits. 

Early dynamical models, in addition, lacked dark matter (DM) halos
\citep{Cretton99,Gebhardt00,Verolme02,Gebhardt03,Shapiro06,Houghton06,Gebhardt07,Nowak08,
  Cappellari09,Siopis09,Gueltekin09,Krajnovic09,vandenBosch10}. Omission
of DM leads to a bias in the masses of the stars and of the central
black hole \citep{Gebhardt09,Rusli13a}. Distributions of stellar
orbits measured without including the effect of DM are
correspondingly unreliable.

For about half of the $\sim$ two dozen galaxies that have been
modelled with DM halos so far, only long-slit {\it Hubble Space
Telescope} (\hst) stellar kinematical data are available
\citep{Shen10,Jardel11,Schulze11}. Such data -- due to the limited
spatial coverage -- do not fully constrain the distribution of central
stellar orbits. Up to now, only three core ellipticals
\citep{Gebhardt11,McConnell12} have reliable measurements of the
central orbital structure using both DM halos and two-dimensional (2D)
kinematic data. Similar measurements for power-law, non-core
elliptical galaxies are yet missing.

Here we present a coherent analysis of the central orbital structure
in 11 massive elliptical galaxies. All of our galaxies have
high-resolution, adaptive-optics assisted, integral field (IFU)
stellar kinematical data obtained with SINFONI at the European Southern Observatory
Very Large Telescope (ESO-VLT). We use
state-of-the-art axisymmetric orbit superposition models
\citep{Schwarzschild79} that include the effects of a central
SMBH, stars and a DM halo.  Six of the
galaxies are core galaxies, while five do not have detectable cores at the
resolution limit of \hst\ \citep{Rusli13a,Rusli13b}. We compare the
orbit distributions in both types of galaxies against each other and
against predictions from core-formation models.

The paper is organized as follows.  In Section~\ref{sec:sample}, we
present an overview of the galaxy sample, observations, and dynamical
models. Anisotropies are described and compared with previous modeling
results in Section~\ref{sec:previous}. In Section~\ref{sec:orbits}, we
compare our results quantitatively with core-formation models. The
paper is summarized in Section~\ref{sec:summary}.

\begin{deluxetable}{lcccccc}
\tablecolumns{7}
\tablewidth{0pt}
\centering
\tablecaption{Summary of the galaxy sample\label{tab:data}}
\tablehead{
\colhead{Galaxy} & \colhead{$\rb$} & \colhead{$\rsoi$} & \colhead{FWHM} & \colhead{$\mbh$} & \colhead{$\mdef$}
 & \colhead{$\sigma_e$}
\\
\colhead{} & \colhead{$(\arcsec)$} & \colhead{$(\arcsec)$} & \colhead{$(\arcsec)$} & \colhead{$(10^9 \, M_\odot)$} & \colhead{$(\mbh)$}
 & \colhead{(km/s)}
}
\startdata
NGC\,1407 & $2.0$ & $2.0$ & $0.19$ & $4.5$ & $0.96$ & $276.1$\\
NGC\,1550 & $1.2$ & $0.7$ & $0.17$ & $3.7$ & $2.95$ & $270.1$\\
NGC\,3091 & $0.6$ & $0.6$ & $0.17$ & $3.6$ & $4.36$ & $297.2$\\
NGC\,4472 & $1.8$ & $1.5$ & $0.47$ & $2.5$ & $1.46$ & $300.2$\\
NGC\,5328 & $0.9$ & $0.7$ & $0.14$ & $4.7$ & $5.70$ & $332.9$\\
NGC\,7619 & $0.5$ & $0.4$ & $0.18$ & $2.5$ & $5.40$ & $292.2$\\
NGC\,1374 & $\ldots$   & $0.8$ & $0.15$ & $0.6$ & $\ldots$ & $166.8$\\
NGC\,4751 & $\ldots$   & $0.4$ & $0.22$ & $1.4$ & $\ldots$ & $355.4$\\
NGC\,6861 & $\ldots$   & $0.4$ & $0.38$ & $2.0$ & $\ldots$ & $388.8$\\
\hline
NGC\,307 & $\ldots$   & $0.2$ & $0.17$ & $0.6$ & $\ldots$ & $201.4$\\
NGC\,1332 & $\ldots$   & $0.4$ & $0.14$ & $1.4$ & $\ldots$ & $292.4$\\
\enddata
\tablecomments{Core radii $\rb$ and mass-deficits $\mdef$ are from
  \citet{Rusli13b}. Black-hole masses $\mbh$, sphere-of-influence
  radii $\rsoi$, the spatial resolution of the kinematic data and
  effective velocity dispersion $\sigma_e$ are from \citet{Rusli13a},
  except for the new results for NGC\,307 and NGC\,1332.\\}
\end{deluxetable}

\section{Observations and Modeling}
\label{sec:sample}
Our galaxy sample is based on the set of high velocity-dispersion
ellipticals in \citet{Rusli13a} where we study the influence of
DM halos on dynamical black hole mass measurements. Details
of the photometric and kinematical data, the data reduction, and our
modeling technique are presented therein. We observed the center of
each galaxy with SINFONI at the ESO-VLT. These very high spatial
resolution data guarantee that the sphere of influence of the central
black hole is well resolved (Table~\ref{tab:data}). We combined these
data with kinematical observations on larger scales, typically out to
about the half-light radius of each galaxy.  Extended spatial coverage
with kinematical data is essential to constrain the distribution of
stellar orbits, in particular to constrain the amount of stars on
radially extended orbits. One galaxy (NGC\,5516) out of the original
10-galaxy sample is omitted here since its kinematical data reach
only out to $r_\mathrm{max} \lesssim 12 \, \rb$ and the orbital
structure in the core is less constrained than in the other galaxies
($r_\mathrm{max} > 20 \, \rb$).  To this sample of nine galaxies we
add two additional high-$\sigma$ galaxies that we also observed with
SINFONI: NGC\,307 and NGC\,1332. The data for NGC\,1332 were already
published in \citet{Rusli11}; the new data for NGC\,307 are described
in Erwin et al. (in preparation). Both new galaxies have been
  analyzed following the same data reduction/modeling scheme as
  described in \citet{Rusli13a}. To constrain the spatial
distribution of the stars we use a combination of \hst\ images,
ground-based images, and our SINFONI data \citep{Rusli13a}.

Six out of the 11 galaxies are core galaxies, while five do not
have a detectable core at the limit of \hst\ resolution
\citep{Rusli13b}. Cores in the observed galaxies were identified and
measured using core-S{\'e}rsic model \citep{Graham03,Trujillo04} fits
to surface brightness profiles that extend out to large radii,
typically more than twice the half-light radius of the galaxy
\citep{Rusli13b}. A summary of the relevant galaxy properties is given
in Table~\ref{tab:data}.

\subsection{Schwarzschild Modeling}
The population of stellar orbits is determined by fitting
Schwarzschild orbit-superposition models to these data. In our models
we assume the mass density $\rho$ of the galaxies to be described by
\begin{equation}
\label{eq:density}
\rho(r) = \mbh \times \delta(r) + \Upsilon \times \nu(r) + \rho_\mathrm{DM}(r),
\end{equation}
where $\mbh$ is the mass of the central SMBH, $\Upsilon$ is the
stellar mass-to-light (M/L) ratio and $\nu$ denotes the deprojected galaxy
luminosity density. The DM density $\rho_\mathrm{DM}$ is
modeled by a logarithmic halo, which is constrained by two parameters
that are included in the fit\footnote{Our models are axisymmetric,
  i.e., the mass density is assumed to vary as a function of radius
  and a polar angle. For the sake of compactness, we omitted the
  angular dependence in Equation~(\ref{eq:density}).}.  Taking into
account the effect of a DM halo is essential to derive
unbiased black hole masses \citep{Gebhardt09,Rusli13a} and, as we will
show in Section~\ref{subsec:dm}, to derive an unbiased orbit
distribution.

To determine the best-fitting mass models, we proceed in two steps. We
first omit the central high-resolution SINFONI data and only vary
$\Upsilon$ and the DM density $\rho_\mathrm{DM}$.  From this first set
of models, we determine $\rho_\mathrm{DM}(\Upsilon)$, the best-fitting
DM density distribution as a function of $\Upsilon$. For the final set
of models, we include the central kinematical constraints from our
SINFONI observations and vary $\mbh$ and $\Upsilon$ and use the
relation $\rho_\mathrm{DM}(\Upsilon)$ determined from the first set of
models \citep{Rusli13a}. In this way, we reduce the originally 4D
parameter space to two smaller ones (3D and 2D, respectively).

In each trial potential, i.e., for each particular choice of $\mbh$,
$\Upsilon$, and $\rho_\mathrm{DM}$, we compute an orbit library
containing about $25\,000$ representative stellar orbits. Half of
these orbits have positive angular momentum $L_z$ along the rotation
axis; the rest have their $L_z$ flipped in the opposite direction. The
contribution of each orbit to the best-fit model in the actual
potential, the so-called orbital occupation number or weight, is
determined by the maximization of
\begin{equation}
\label{eq:maxs}
S-\alpha \chi^2,
\end{equation} 
subject to the constraints imposed by the deprojected galaxy
luminosity density $\nu$. In Equation~(\ref{eq:maxs}), $S$ denotes the
Boltzmann entropy and $\chi^2$ quantifies the deviations between the
observed line-of-sight velocity distributions and the model. The
entropy maximization and $\chi^2$ minimization is done iteratively for
$\sim 30$ different values of the regularization parameter
$\alpha$. We start with $\alpha=0$ and stop when an increase in
$\alpha$ does not reduce the $\chi^2$ anymore. Models with different
$\alpha$ reproduce the density constraints equally well but differ in
their internal orbital structure: at $\alpha=0$, the entropy is
maximized, while for larger $\alpha$, the orbital weights are
adjusted more to fit the model to the kinematical observations (as good as
the actual trial potential allows). The results presented in this
paper are based on models with maximum $\alpha$, i.e., the orbit
distribution is entirely determined by the observational data. A
detailed description of our implementation of the Schwarzschild method
can be found in \citet{Thomas04,Thomas05}.

\begin{figure*}
\centering
\includegraphics[scale=0.77]{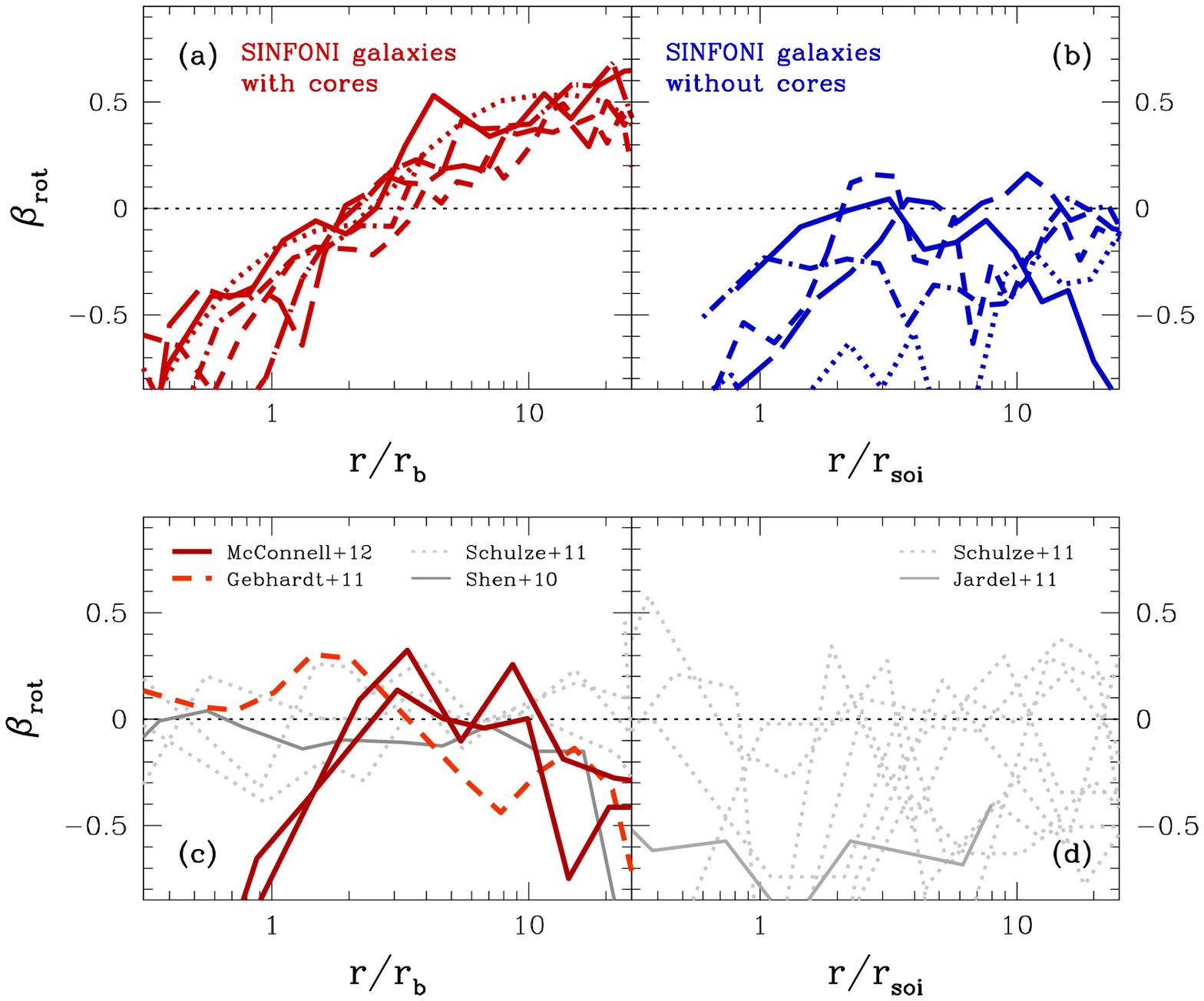}
\caption[]{Central anisotropy profiles $\betarot$ of early-type
  galaxies from models including a central SMBH, stars and a DM
  halo. Top: SINFONI galaxies with cores (panel (a); radii are scaled
  by the core radius $\rb$) and SINFONI galaxies without cores (panel
  (b); radii are scaled by the sphere of influence $\rsoi$).  Our
  models are based on IFU kinematics.  Bottom: previous galaxy models
  for core galaxies (c) and non-core galaxies (d). We use the
  core/non-core classification of \citet{Rusli13b}.  Results based on
  IFU data similar to ours are highlighted
  \citep{Gebhardt11,McConnell12}. All galaxy models assume axial
  symmetry.
  \label{fig:dm}}
\end{figure*}

\section{Stellar orbits in the centers\\ of massive ellipticals}
\label{sec:previous}
Figures~\ref{fig:dm}(a) and (b) show the distribution of stellar
orbits in our galaxies as measured from the best-fit models.  The
figure shows for each galaxy the radial profile of the orbital
anisotropy parameter:
\begin{equation}
\label{eq:betarot}
\betarot = 1 - \frac{\sigma_\mathrm{rot}^2}{\sigma_r^2},
\end{equation}
where 
\begin{equation}
\label{eq:sigrot}
\sigma_\mathrm{rot}^2 = (\sigma^2_\vartheta + \sigma^2_\varphi + v_\varphi^2
)/2
\end{equation} 
is the light-weighted spherically averaged second velocity moment in
the tangential direction and $\sigma_r$ the corresponding moment in
the radial direction.  In Figure~\ref{fig:dm}(a), the radial
coordinate is scaled by the core radius $\rb$ and only core galaxies
are shown. In Figure~\ref{fig:dm}(b), only the non-core galaxies are
shown and the radial coordinate is scaled by the sphere-of-influence
radius $\rsoi = G\,\mbh/\sigma_e^2$.  In our core galaxies $\rsoi \sim
\rb$ (Table~\ref{tab:data}).

\subsection{Homology in the Velocity Dispersion Tensor}
The dynamical structure around the SMBHs in core galaxies is different
from that in non-core, power-law galaxies.
The central anisotropy profiles of the six core galaxies are
remarkably uniform. Stellar motions inside the core radius $\rb$ are
strongly dominated by tangential orbits ($\betarot<0$), while outside
$\rb$ radial orbits quickly take over ($\betarot>0$). In fact, there is a
strong homology in the anisotropy profiles: when scaled by the core
radius $r_b$, the radial profiles of the classical anisotropy
parameter $\betarot(r)$ are nearly identical in core galaxies.  The five
galaxies without cores do not show any coherent anisotropy pattern,
but instead are tangentially biased on all spatial scales.

The homogeneity of the stellar orbits is not directly enforced by the
break in the light profiles. To check this, we also built
maximum-entropy models \citep[e.g.][]{Thomas09} for each core
galaxy. These models reproduce the light profile equally well but are
isotropic in the center. They do not fit the observed kinematics,
however, having higher central velocity dispersions and steeper
velocity dispersion gradients than observed.

\subsection{Previous Anisotropy Measurements}
A compilation of previous anisotropy measurements in the centers of
early-type galaxies is shown in Figures~\ref{fig:dm}(c) and (d). Models
based on IFU data are highlighted and referenced on the left, while
models fitted to long-slit data are quoted on the right. All models
shown in Figure~\ref{fig:dm} include a SMBH, stars, and a DM halo.

The core-galaxy models of \citet{Schulze11} do not show a tendency
toward tangential anisotropy inside the core radius $\rb$: instead,
all three core galaxies in their sample (NGC\,3608, NGC\,4291, and
NGC\,4649) show nearly constant and isotropic dispersion profiles
(Figure~\ref{fig:dm}(c)).  The recent models of the two brightest
cluster galaxies NGC\,3842 and NGC\,7768
\citep{McConnell12}\footnote{For our comparison, we leave out two of
  the four galaxies from \citet{McConnell12}: NGC\,2832, because only
  an upper limit for the mass of the central SMBH was derived, and
  NGC\,4889, because its kinematical data reach only out to $r \la
  13\,\rb$.} do show increasing tangential anisotropy, as do our
models.  We attribute the different results to the different data that
were used: \citet{Schulze11} use long-slit \hst\ data that do not
fully cover the central galaxy regions and leave most stellar orbits
unconstrained. In contrast, the study of \citet{McConnell12} and ours
are based on IFU spectroscopic data in the center that provide full
spatial coverage and the most complete constraints on stellar
orbits. This indicates that IFU data are important for understanding
the central orbital structure in core galaxies. However, we note that
the latest model of M87 by \citet{Gebhardt11} is based on IFU data as
well. It shows tangential anisotropy only inside $r \la 0.2 \, \rb$.
A direct comparison to our work is hampered by the fact that
\citet{Gebhardt11} do not fit a stellar M/L ratio but use the
dynamical stellar M/L from \citet{Murphy11}, which was determined from
kinematical data with lower central resolution.  In addition to this
difference in the modeling approach, the velocity dispersion profile
of M87 rises steeply inside the core region, while our core galaxies
have nearly constant central velocity dispersion profiles
(cf. \citealt{Gebhardt11} and \citealt{Rusli13a}).

The situation for power-law galaxies is less clear due to the complete
lack of comparison galaxies studied based on IFU data and models
similar to ours: the existing models with DM for non-core galaxies by
\citet{Schulze11} and the ones for NGC\,4594 \citep{Jardel11} are
based on long-slit data. Our $\betarot$ seem slightly less negative
than the ones in Figure~\ref{fig:dm}(d). Whether this is due to the
use of IFU versus long-slit data is unclear. The published
anisotropies of \citet{Schulze11} are along the galaxies' major axes,
where the influence of $v_\varphi$ on the anisotropy is largest. This
could also explain their more extreme tangential anisotropies, apart
from intrinsic variations in the orbital structure among non-core
galaxies, in which stellar orbits seem to be populated in a less
coherent manner than in core galaxies.

\subsection{Core Rotation}
\begin{figure*}
\centering
  \includegraphics[scale=0.77]{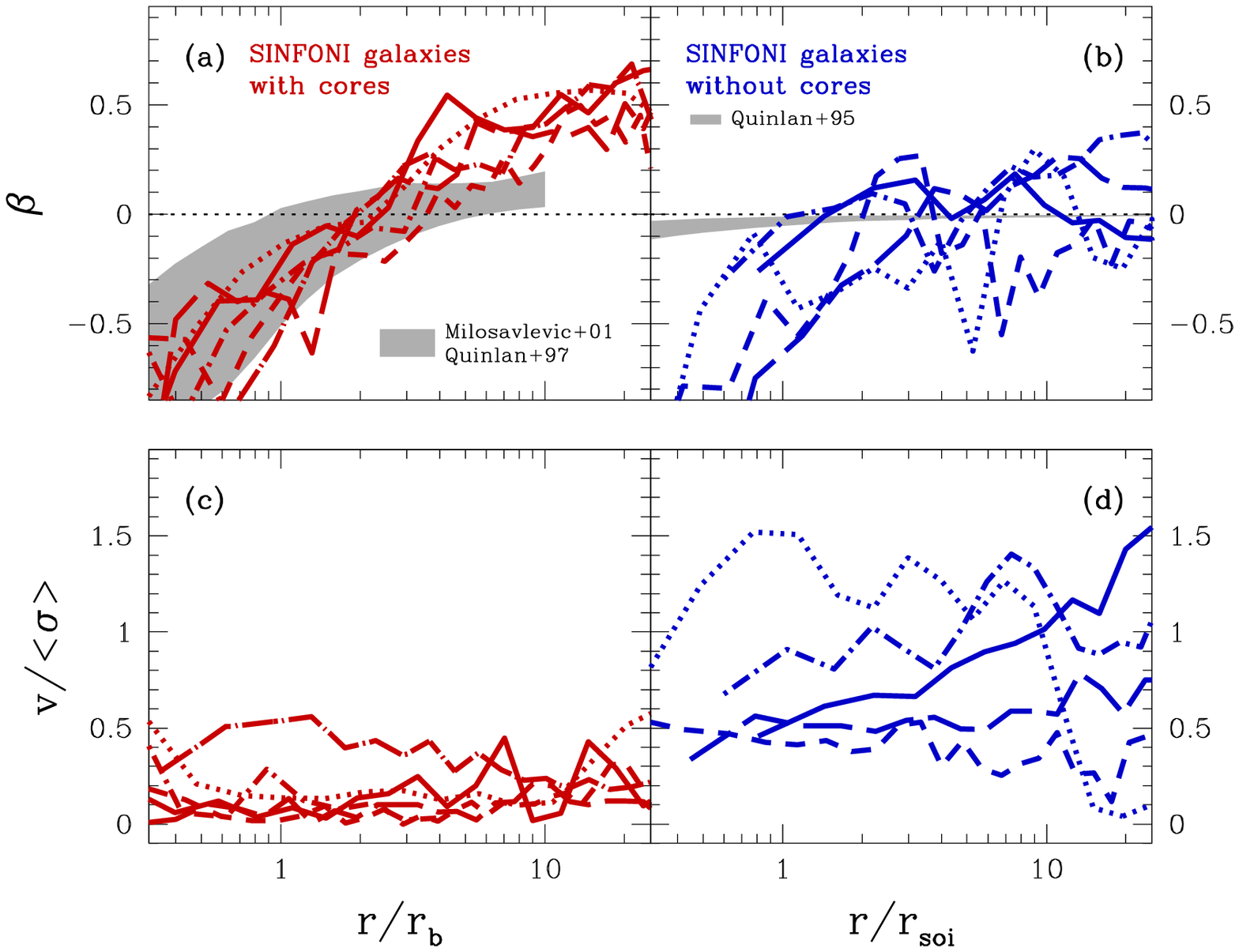}
  \caption[]{Top: intrinsic anisotropy profiles $\beta(r)$ of the
    SINFONI galaxies with cores (a) and without cores (b). Shaded
    areas show predictions from numerical $N$-body simulations of
    core scouring by equal-mass binary black holes (panel (a):
    \citealt{Quinlan97,Milosavljevic01}) and from adiabatic
    SMBH growth models (panel (b): \citealt{Quinlan95}). Bottom:
    rotation profiles along the major axis in the six core galaxies
    (c) and the five galaxies without cores (d). Rotation velocities
    are scaled by the local velocity dispersion $\langle \sigma
    \rangle^2 \equiv (\sigma_r^2 + \sigma_\vartheta^2 +
    \sigma_\varphi^2)/3$.}
\label{fig:intrinsic}
\end{figure*}

Figure~\ref{fig:intrinsic} shows the central orbital structure of our SINFONI
galaxies, split up into the {\it intrinsic} velocity dispersion anisotropy
\begin{equation}
\label{eq:beta}
\beta \equiv 1 - \frac{\sigma_\mathrm{tang}^2}{\sigma_r^2},
\end{equation}
with
\begin{equation}
\label{eq:sigtang}
\sigma_\mathrm{tang}^2 \equiv (\sigma_\vartheta^2 + \sigma_\varphi^2)/2
\end{equation}
(panels (a) and (b)) and the stellar rotation component (panels (c)
and (d)).  The six core galaxies show little to no rotation in their
central regions, but all five power-law galaxies rotate with values of
$\sim 0.5 - 1.5$ times the local velocity dispersion $\langle \sigma
\rangle$. Hence, in core galaxies, $\betarot \sim \beta$ and the
similarity in the central orbital structure is due to a homology in
the intrinsic velocity dispersion tensor. In power-law galaxies,
$\betarot < \beta$ and a significant fraction of the tangential
anisotropy ($\Delta \beta \sim 0.2 - 0.4$) comes from stars that carry
net angular momentum. The remaining anisotropy in the velocity
dispersion tensor is more similar to that of core galaxies.  Yet, even
after separating out the rotating component, power-law galaxies still
have a slightly more tangential orbit distribution.

\subsection{The Importance of Dark Matter Halos}
\label{subsec:dm}
\begin{figure*}
\centering
\includegraphics[scale=0.77]{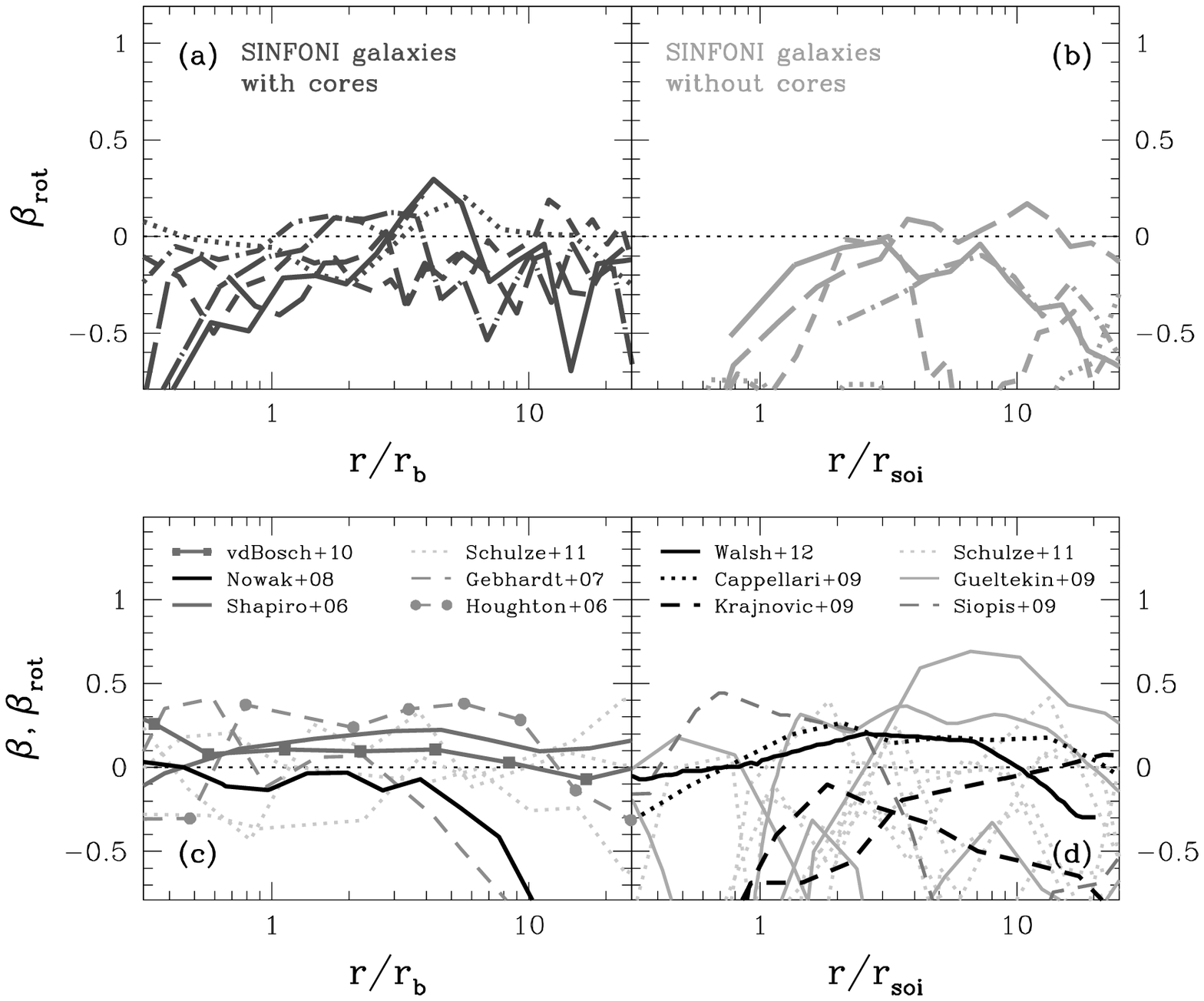}
\caption[]{As in Figure~\ref{fig:dm}, but for models that only include
  a SMBH and stars ({\it no} DM halo). Top: SINFONI galaxies; bottom:
  previous models. Results based on IFU data are again highlighted
  \citep{Shapiro06,Nowak08,Cappellari09,Krajnovic09,vandenBosch10,Walsh12}.
  Core galaxies are on the left ((a) and (c)); power-law galaxies are
  on the right ((b) and (d)).  All galaxy models are axisymmetric,
  except for NGC\,3379 and NGC\,3998 (triaxial;
  \citealt{vandenBosch10,Walsh12}) and NGC\,1399 (spherical;
  \citealt{Houghton06}). Note that \citet{Gebhardt07},
  \citet{Gueltekin09}, and \citet{Schulze11} used $\betarot$, while
  all other results shown in panels (c) and (d) are for $\beta$
  (details in the text).\label{fig:nodm}}
\end{figure*}

Figure~\ref{fig:nodm} is similar to Figure~\ref{fig:dm}, but shows the
distribution of stellar orbits from models that only include a SMBH
and stars ({\it no} DM halo). Ignoring the contribution of DM
halos in the models changes the derived orbital structure
significantly. In fact, in our core-galaxy models without DM
(Figure~\ref{fig:nodm}(a)), the homology (as seen in Figure~\ref{fig:dm}(a)) 
disappears. In the outer parts, the omission of
DM reduces the amount of stars on radial orbits, which is a
reflection of the well-known mass-anisotropy degeneracy, i.e., that
missing mass can be partly compensated for by a more tangential orbit
distribution \citep[e.g.][]{Binney82,Gerhard93}.

\citet{Schulze11} did a similar comparison between models that
included a SMBH, stars, and DM and models with only a SMBH and
stars. Their results change only a little when DM is included. We
attribute the difference to our models, which change significantly
when the effect of a DM halo is included, to the fact that
\citet{Schulze11} do not allow for as much freedom in the distribution
of DM as we do in our models (cf. \citealt{Rusli13a}).  In addition,
the models of \citet{Schulze11} rely on {\it long-slit} \hst\ data,
which, by their one-dimensional (1D) nature, do not fully cover the
central galaxy regions and leave most of the stellar orbits
unconstrained (see above). Our models presented here are instead based
on {\it IFU} spectroscopic data that fully cover the galaxy centers
and provide the most complete constraint on stellar orbits.

The bottom row of Figure~\ref{fig:nodm} compiles previous anisotropy
measurements, where those based on IFU data are again highlighted and
referenced on the left, while models fitted to long-slit data are
quoted on the right. Note that \citet{Gebhardt07}, \citet{Gueltekin09},
and \citet{Schulze11} used $\betarot$, while all other results shown
in Figures~\ref{fig:nodm}(c) and (d) are for $\beta$.

Core galaxy models without DM generally yield isotropic or even
slightly radial orbit distributions ($\betarot,\, \beta \ga 0$;
Figure~\ref{fig:nodm}(c)). According to Equation~(\ref{eq:sigrot}),
$\betarot$ is always lower than $\beta$. Thus, different definitions
of $\beta$ cannot explain why many previous studies found isotropic or
slightly radial orbit distributions in core galaxies (around $\rb$).
But the inclusion of rotation can partly explain the difference among
the non-core galaxies in Figure~\ref{fig:nodm}(d).  In any case, even with
models that neglect DM halos, stellar orbits in power-law galaxies
appear more strongly biased in the tangential direction than do those
in core galaxies.

From Figure~\ref{fig:dm}, it becomes clear that models that do include
DM and a SMBH recover different orbit distributions depending on
whether they are based on 2D spectroscopic data or 1D long-slit
data. Without a DM halo, instead, the difference between models fit to
IFU data and models fit to long-slit data is less significant
(Figure~\ref{fig:nodm}); if anything, models fit using IFU data are
slightly more radially anisotropic. This indicates that IFU data are
important for the recovery of the orbital structure, because they help
to constrain the gravitational potential, in particular the mass of
the central SMBH more tightly. In \citet{Rusli13a}, we find that SMBH
masses change on average by a factor of $\sim 2$ when including a DM
halo; in a similar study, although based on 1D \hst\ data,
\citet{Schulze11} find an increase in SMBH masses of only $\sim 20$\%.

We conclude that a reliable reconstruction of the central orbital
structure requires the {\it combination} of IFU data and DM halos in
the models; neither factor by itself is sufficient.

\subsection{Uncertainties}
Almost all previous works -- and ours -- assume axial symmetry for the
studied galaxies. The only published triaxial models, for the
  centers of NGC\,3379 and NGC\,3998 \citep{vandenBosch10,Walsh12},
  are dominated by box orbits in the innermost regions. Box orbits do
  not exist in axisymmetric potentials. However, the associated radial
  anisotropies in the triaxial models of NGC\,3379 and NGC\,3998 are
  similar to those in other galaxies derived from similar data and
  assumptions about the mass profile (no DM halo). Moreover, the
  triaxial model of NGC\,3379 is similar to the axisymmetric model of
  NGC\,3379 by \citet{Shapiro06} outside $r \ga \rb/2$. NGC\,3379 and
  NGC\,3998 thus do not provide evidence that the assumption of axial
  symmetry induces a strong bias.  This could be expected for the
  central regions where the potential is dominated by the SMBH, i.e.,
  spherical and Keplerian. Final conclusions can only be drawn when
triaxial models with a DM halo are available.

The orbit distributions derived for the centers of power-law galaxies
are probably more uncertain than those for core galaxies. At the
innermost resolved data point there are still orbits with pericenters
at even smaller radii. These orbits are only partly constrained, by
light scattered from the center to larger radii through point-spread
function effects. Because power-law galaxies have brighter centers,
constraints from presently unresolved spatial regions are likely to
affect dynamical models more strongly than in core galaxies with their
faint inner regions.

\section{Core formation}
\label{sec:orbits}

The remarkable homogeneity of the orbital structure in the core
galaxies and the fact that the transition from tangential to radial
anisotropy is strongly correlated with the core radius $\rb$ provide
an observational link between core formation and the population of
central stellar orbits. Moreover, the core formation process must be
very uniform to produce the observed homology in the distribution 
of core stellar orbits.

\subsection{Adiabatic Growth and Core Expansion}
Tangential anisotropy in the core could be the result of adiabatic
black hole growth \citep{Goodman84}. For two reasons, however, this
process is unimportant for stellar orbits in core galaxies.  First,
the effect of adiabatic black hole growth on the orbital structure is
weak and confined to the region inside $\rsoi$: the shaded area in
Figure~\ref{fig:intrinsic}(b) shows models of \citet{Quinlan95}.  They
start from an isotropic orbit distribution and initial central
logarithmic density slopes of $\gamma=\{0,-1\}$. Even when the final
black hole mass is up to 10\% of the galaxy mass, the change in
the central anisotropy remains modest ($| \Delta \beta| \le 0.3$; see
also \citealt{Young80} and \citealt{Goodman84}). If adiabatic
black hole growth was the main driver behind the observed anisotropy
in the cores of massive ellipticals, then the orbits would be required
to be highly coherent and correlated with $\rsoi$ already before the
black hole growth. Our coreless galaxies do not provide evidence for
this. Second, adiabatic black hole growth steepens the central light
profile \citep{Young80} and thus cannot explain the formation of cores
in the first place. Core creation through dynamical friction-driven
angular momentum transfer from infalling objects \citep{Goerdt10} and
core expansion in reaction to gas ejection by an active galactic
nucleus \citep{Martizzi12} are unlikely as well. These processes are
not directly linked to the mass of the central black hole and hence
cannot explain the fact that cores sizes correlate tightly with $\mbh$
\citep{Rusli13b}.  These processes might, however, be important for
less massive, fast-rotating core galaxies \citep[e.g.][]{Krajnovic13}.

\subsection{Core Formation by Binary Black Holes}
\citet{Quinlan97} studied the response of a galaxy's orbital
anisotropy and its 3D density profile to a central, equal-mass black
hole binary, starting from isotropic systems with various inner
logarithmic density slopes. We fit core radii $\rb$ to their
simulations with black hole masses in the range $\mbh = (0.005 -
0.02) \, M_\mathrm{tot}$, and initial slopes $\gamma = \{-1,-2\}$,
typical for power-law ellipticals.  To ensure compatibility with the
core radii measured in real galaxies, the fits were performed using
the (approximate) deprojection of the core-S´ersic model given by
\citet{Terzic05}.  When scaled by these core radii, the anisotropy
profiles of the \citet{Quinlan97} simulations are all very
similar. They define the lower boundary of the shaded region in
Figure~\ref{fig:intrinsic}(a).

\citet{Milosavljevic01} investigated the entire merging process of two
identical elliptical galaxies with initial central density slopes
$\gamma = -2$ and also measured core radii from the remnant's final
projected surface mass profile. Their definition of $\rb$ (radius of
maximum curvature) differs slightly from our $\rb$ derived with
core-S´ersic models, but the difference is small ($\sim 10$\%). The
results of \citet{Milosavljevic01} define the upper boundary of the
shaded area in Figure~\ref{fig:intrinsic}(a).

Both simulations predict a transition in the orbital structure around
the core radius: inside $\rb$, tangential orbits dominate; outside
the core, radial orbits take over. Similar anisotropy patterns have
been derived in other simulations as well
\citep{Zier01,Meiron10,Antonini12}.

The observed orbital structure of the core galaxies follows
this prediction closely, while non-core galaxies do not
exhibit the characteristic change in the orbital structure from
tangential to radial anisotropy. This provides the first
convincing dynamical evidence for core scouring in massive
elliptical galaxies.

\subsection{Mass Deficits and Merging Histories}
A single binary is expected to eject about its own mass in stars from
the center \citep{Merritt06}. Larger mass deficits can result from:
\begin{enumerate}
\item the cumulative effect of repeated mergers \citep{Merritt06}
\item repeated core passes of the post-binary-merger black hole due to
  gravitational-wave recoil \citep{Gualandris08} ($\mdef \lesssim 5 \,
  \mbh$)
\item multiple black hole systems that might form at high
  redshifts \citep{Kulkarni12} ($\mdef \lesssim 5 \, \mbh$).
\end{enumerate}
The mass deficits derived for our core galaxies are $\mdef \sim 1-6 \,
\mbh$ (Table~\ref{tab:data}) and, again, fit the numerical
predictions well.

Most of our galaxies have mass deficits consistent with only one major
merger, in agreement with their orbital anisotropy. A corresponding
number of minor mergers, however, would give similar mass
deficits. Studies of test-particle orbits around equal-mass as well as
unequal-mass black hole binaries suggest that the central anisotropy,
like the mass deficit, does not carry specific information about the
merging history \citep{Meiron10}.

Such information could, however, be encoded in the central rotation
velocities. In equal-mass merger simulations, rotation velocities
similar to the local velocity dispersion are found
\citep{Milosavljevic01,Meiron13}. Unequal-mass binaries likely induce
less rotation through dynamical friction \citep{Meiron13}.  Rotation
velocities in our observed cores are small
(Figure~\ref{fig:intrinsic}). This might favor a sequence of several
minor mergers, where, if the merger orbits are random, each new merger
would tend to wash out the rotation signal left over from the previous
merger. The stochastic behavior of possible multiple black hole
systems, formed as a consequence of successive minor mergers, might
further reduce core rotation in real galaxies. Constraints from galaxy
scaling relations, however, limit the mass ratios of precursor
galaxies as well \citep{KormendyBender13}.  To put more quantitative
constraints on the detailed formation histories of core galaxies, we
need (1) numerical simulations that explore a wider range of initial
conditions and (2) discussion of stellar population properties
(something we plan for a forthcoming paper).

\subsection{Non-core Galaxies}

The five non-core galaxies of our sample are not consistent with
simple models of an adiabatic black hole growth into an originally
isotropic stellar distribution: the relative overpopulation of
tangential orbits with stars is stronger and spread over a spatially
more extended region than expected from these models. 

They are, however, compatible with the idea that these galaxies form
in gas-rich mergers, where any pre-existing core is ``refilled'' by
gas that falls in during/after the merger, settles into a rotating
configuration, and then forms new stars. In fact, the intrinsic
anisotropy (without rotation) of non-core galaxies shows a slight
tendency toward a gradient from tangential anisotropy in the center
toward a more isotropic or radially dominated orbit system at $r \ga
\rsoi$ that is similar to the transition in core galaxies, although with
significantly more scatter. This imprint in the stellar orbits could
be the relic of a transient core structure that occurred at some point
during the formation of these galaxies and led to a partially depleted
core (although the dynamical friction on the SMBHs is much stronger in
a gaseous environment and the timescale for interactions between
stars and the SMBHs is correspondingly short). In this scenario, a
strong rotating component would also be present, consisting of the
stars that formed later, refilling the core and possibly producing
the central extra light component discussed by, e.g., \citet{Cote06},
\citet{KormendyBender09}, \citet{KFCB} and \citet{KormendyHo13}.

Another explanation for the central tangential orbits might be a
reshaping of the stellar orbits similar to the adiabatic SMBH-growth
scenario, but originating from the formation of a concentrated stellar
density cusp during a dissipational merger.  Quantitative predictions
for the central orbital structure from numerical simulations including
gas dissipation and star formation are not available yet, however.

\section{Summary}
\label{sec:summary}
We analyze the central distribution of stellar orbits in 11
massive early-type galaxies, derived from orbit-based Schwarzschild
models including a central SMBH, stars, and DM halos. Six of
our galaxies are core galaxies; the other five have power-law
centers. This is the first consistent analysis of a sample of massive
core as well as non-core early-type galaxies to study the homology of
stellar orbits based on homogeneous, high-resolution, IFU
spectroscopic data and state-of-the-art modeling techniques.

The comparison of previous modeling attempts using either data with
less spatial coverage and/or models that lacked DM halos
indicates that both 2D IFU data coverage in the center and accounting
for the effects of DM are necessary for unbiased measurements
of the distribution of stellar orbits.

In core galaxies, the distribution of stellar orbits shows a coherent
change from mostly tangential orbits inside $\rb$ to mostly radial
orbits further out. Moreover, there is a remarkable homology: when
scaled by the core radius $\rb$, the radial profiles of the classical
anisotropy parameter $\beta(r)$ are nearly identical in core galaxies.
Power-law galaxies do not take part in this homology. The scatter in
the distribution of stellar orbits in these galaxies is larger, but
indicates a preference of tangential orbits as well.

The orbital structure observed in the centers of core galaxies matches
the predictions of black hole binary models. They do not match
predictions of adiabatic growth models, although both models favor
tangential orbits in the center.  The difference is (1) in the
magnitude of the tangential bias and (2) in the change of the orbit
distribution with radius, in particular that radial orbits dominate
outside the core in SMBH-binary models. Our observations provide the
first convincing direct dynamical evidence for core scouring in
massive elliptical galaxies.

\section*{Acknowledgements}
Some of the data used in this paper were obtained using SINFONI at the
Very Large Telescope (VLT) and from the Mikulski Archive for Space
Telescopes (MAST). The VLT is operated by the European Southern
Observatory on Cerro Paranal in the Atacama desert of northern
Chile. STScI is operated by the Association of Universities for
Research in Astronomy, Inc., under NASA contract NAS5-26555.  This
paper also includes data taken at The McDonald Observatory of The
University of Texas at Austin.

R.P.S. and R.B. acknowledge support from the DFG Cluster of Excellence Origin and
Structure of the Universe.  P.E. was supported by the Deutsche
Forschungsgemeinschaft through the Priority Programme 1177 ``Galaxy
Evolution''.

Finally, this work has made use of the NASA/IPAC Extragalactic
Database (NED), which is operated by the Jet Propulsion Laboratory,
California Institute of Technology, under contract with the National
Aeronautics and Space Administration.


\end{document}